\documentclass[10pt,conference]{IEEEtran}

\usepackage{balance,booktabs,cite,tabularx,xspace,xurl}
\usepackage[T1]{fontenc}
\usepackage{microtype}
\usepackage{tikz}
\usetikzlibrary{arrows.meta,positioning}
\usepackage[hidelinks]{hyperref}

\newcommand{\bench}{\textsc{AIFaultBench}\xspace}

\newcommand{\videoURL}{\url{https://youtu.be/PHK_g7sa-0g}}

\begin{document}

\title{AIFaultBench: A Reproducible Benchmark of Real-World Faults in AI Software Systems}

\author{
\IEEEauthorblockN{Mehil B Shah}
\IEEEauthorblockA{
\textit{Dalhousie University}\\
Halifax, Canada\\
shahmehil@dal.ca}
\and
\IEEEauthorblockN{Mohammad Masudur Rahman}
\IEEEauthorblockA{
\textit{Dalhousie University}\\
Halifax, Canada\\
masud.rahman@dal.ca}
\and
\IEEEauthorblockN{Foutse Khomh}
\IEEEauthorblockA{
\textit{Polytechnique Montr\'eal}\\
Montreal, Canada\\
foutse.khomh@polymtl.ca}
}

\maketitle

\begin{abstract}
AI-based software systems remain difficult to debug because their failures can involve interactions among code, models, data, dependencies, hardware, and execution environments. Existing benchmarks cover specific classes of AI systems, while concerns about fault realism and reproducibility highlight the need for documented execution conditions and evidence. We present \bench, a benchmark of 770 real-world bugs from 105 AI-based software repositories across 76 organizations. It covers six classes of AI systems: machine learning, deep learning, reinforcement learning, systems based on large language models, agentic AI, and AI tooling and infrastructure. Each bug record connects the original report and source revision to structured metadata, environment setup scripts, a reproduction trigger, and investigation evidence. We reproduced 652 bugs (84.7\%) under the recorded execution conditions and documented the outcomes of the remaining 118 attempts. Through its broad coverage, documented reproduction evidence, and common structure, \bench supports empirical studies of AI faults, debugging tool evaluation, practical failure investigation, and teaching with real-world examples. Dataset: \url{https://zenodo.org/records/21782307}. Video: \videoURL.
\end{abstract}

    
\section{High-Level Overview}

AI-based software systems combine conventional software components with learned models, data processing pipelines, and execution environments. Their failures can therefore span source code, models, data, dependencies, and hardware. Cutting-edge AI-based systems such as large language models (LLMs) and agentic systems further involve very large-scale models, external tools, and memory, making their bugs even more complex~\cite{shah2026agentic}. Over the last few years, the software engineering community has developed several methods to debug these AI systems~\cite{jahan2025improved,shah2026imitation}. To effectively evaluate these debugging methods, we need not only the affected version of an AI-based system but also the execution conditions that trigger the failures. In other words, an evaluation benchmark should provide the artifacts needed to reproduce AI bugs, making benchmark construction a challenging task.

Existing benchmarks of software bugs serve as valuable resources for developers and researchers. Defects4J~\cite{just2014defects4j}, BugsInPy~\cite{widyasari2020bugsinpy}, and SWE-bench~\cite{jimenez2024swebench} provide hundreds of real-world bugs, but they are mostly collected from conventional software projects written in Java or Python. Recent benchmarks such as Denchmark~\cite{kim2021denchmark}, Defects4ML~\cite{morovati2023faultload}, and gDefects4DL~\cite{liang2022gdefects4dl} capture faults from machine learning and deep learning systems. Although these AI-specific benchmarks provide useful datasets and relevant artifacts, they target specific classes of systems, such as machine learning or deep learning systems. A recent study~\cite{jahangirova2024real} reports that only 18.5\% of the examined faults from the existing benchmarks (e.g., Defects4ML) met its realism criteria (i.e., the presence of original buggy code, fixes, and training data) and only 52\% of them were reproducible. Given these observations, a benchmark that curates faults and failures from multiple classes of AI-based systems and connects each fault to relevant artifacts (e.g., source code, dependencies, and reproduction scripts) and execution evidence can greatly benefit the evaluation of debugging methods.

\begin{table}[t]
\caption{Composition of \bench.}
\label{tab:overview}
\centering
\setlength{\tabcolsep}{3pt}
\renewcommand{\arraystretch}{1.12}
\setlength{\aboverulesep}{0.35ex}
\setlength{\belowrulesep}{0.45ex}
\begin{tabularx}{\columnwidth}{@{}Xrr@{}}
\toprule
\textbf{Class of AI systems} & \textbf{Faults} & \textbf{Share} \\
\midrule
Deep learning & 307 & 39.9\% \\
Large language models & 151 & 19.6\% \\
Agentic AI & 130 & 16.9\% \\
Machine learning & 118 & 15.3\% \\
Reinforcement learning & 40 & 5.2\% \\
Tooling and infrastructure & 24 & 3.1\% \\
\bottomrule
\end{tabularx}
\vspace{-2em}
\end{table}

We present \bench, a benchmark of 770 real-world faults from 105 AI-based software repositories maintained by 76 organizations~\cite{aifaultbench}. These faults were submitted as bug reports between July 2016 and July 2026 and stem from six classes of AI systems, as summarized in Table~\ref{tab:overview}. In the benchmark, each bug report is connected to the target version of its codebase, dependency specifications, setup scripts, a reproduction trigger, and investigation evidence. We capture the affected version of the codebase based on its public repository and recorded commit, preserving source provenance, and attempt to reproduce each bug. We reproduced 652 faults (84.7\%) under the recorded execution conditions and documented why the remaining 118 faults were not reproduced. We spent $\approx$1,640 person-hours to construct the benchmark.

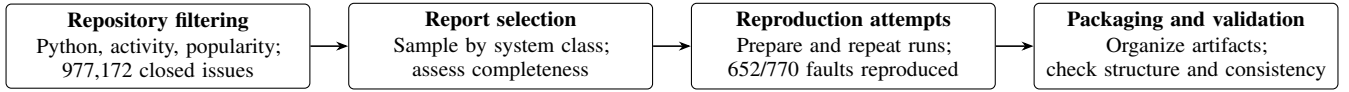
\begin{figure*}[t]
\centering
\begin{tikzpicture}[
    node distance=5mm,
    stage/.style={
        draw,
        rounded corners=2pt,
        align=center,
        text width=0.21\textwidth,
        inner xsep=3pt,
        inner ysep=3pt,
        font=\footnotesize
    },
    flow/.style={
        -{Stealth[length=1.6mm]},
        semithick
    }
]

\node[stage] (discovery) {
    \textbf{Repository filtering}\\[1pt]
    Python, activity, popularity;\\
    977,172 closed issues
};

\node[stage, right=of discovery] (selection) {
    \textbf{Report selection}\\[1pt]
    Sample by system class;\\
    assess completeness
};

\node[stage, right=of selection] (reproduction) {
    \textbf{Reproduction attempts}\\[1pt]
    Prepare and repeat runs;\\
    652/770 faults reproduced
};

\node[stage, right=of reproduction] (validation) {
    \textbf{Packaging and validation}\\[1pt]
    Organize artifacts;\\
    check structure and consistency
};

\draw[flow] (discovery) -- (selection);
\draw[flow] (selection) -- (reproduction);
\draw[flow] (reproduction) -- (validation);

\end{tikzpicture}
\caption{Construction and validation of \bench.}
\label{fig:collection-process}
\end{figure*}

Together, \bench offers three key benefits over existing benchmarks. First, it has \textbf{broad fault coverage} across six classes of AI systems, supporting a comprehensive evaluation of the debugging methods. Second, it supports \textbf{fault reproducibility} through documented reproduction steps and execution results, equipping researchers with valuable insights and an oracle during the evaluation of their debugging methods. The benchmark also distinguishes successfully reproduced faults from faults that were not reproduced and documents the outcomes. Finally, it supports \textbf{reusability} through an index and bundle structure and allows the same bug reports and artifacts to be used across experiments. Researchers can reuse these artifacts to evaluate their methods targeting fault reproduction, localization, testing, and repair in various AI-based systems.

\section{Dataset Construction and Validation}
Fig. 1 shows the workflow of our benchmark dataset construction and subsequent validation. We discuss its different steps as follows.

\subsection{Data Collection and Provenance} First, we selected AI-based repositories and issue reports from GitHub using repository topic tags and common repository selection criteria adopted from existing work~\cite{shah2026imitation,morovati2023faultload,morovati2024common,shah2026agentic}. We then retained repositories with activity (e.g., commits, merged pull requests, or issue updates) since January 2023 and at least 500 stars. After removing duplicates, we obtained 5,365 repositories. Requiring each repository to have at least 10 watchers reduced this collection to 4,038 repositories. We then retained 3,289 repositories with at least 10 closed issues, yielding 977,172 closed issues for subsequent selection.

Before performing further analysis, we defined six system classes in Table~\ref{tab:overview} to cover established and emerging AI approaches and their supporting infrastructure, motivated by trends in AI research and adoption~\cite{maslej2025artificial}. From the initial pool of 977,172 closed issues, we sampled 1,000 candidate bug reports across the six system classes, using a practical sample size for manual analysis. We assessed their quality using information-completeness criteria adapted from prior work~\cite{bettenburg2008makes,morovati2023faultload,shah2026imitation}. According to the criteria, a bug report should include a description of the failure, reproduction steps or executable examples, diagnostic evidence such as error messages or stack traces, and information about dependencies and required data. We examined the candidate bug reports and associated discussions for this information, assigning $+1$ for each satisfied criterion and $-1$ for each unsatisfied criterion. We retained reports with a positive total score and used the corresponding fix commits as evidence that developers had investigated the reported bugs. After applying these criteria and excluding false positives (e.g., support questions, feature requests), we retained 770 bug reports from 105 AI-based systems for our benchmark dataset.

Besides a bug report, we collected its URL, affected codebase, timestamps, labels, reactions, discussions, and links to fixes. The URL, repository, and revision identify each bug's origin, whereas the discussions preserve its context. We also assign each bug a unique identifier that connects these records, enabling users to navigate the dataset more easily.

\subsection{Bug Reproduction and Evidence Collection} We attempted to manually reproduce each selected bug using a duration threshold of three hours (Fig.~\ref{fig:collection-process}). We manually created the reproduction scripts using the edit actions and information requirements identified in a prior study~\cite{shah2025towards} and the reproduction workflow developed in RepGen~\cite{shah2026imitation}. These findings guided us to identify the dependencies, required inputs, and operations needed to reproduce the reported failure.

We executed the above scripts and compared the observed behaviour with the behaviour described in the original bug report. To reduce nondeterminism, we fixed random seeds and relevant execution settings, using the reproduction guidelines from existing work~\cite{chen2022towards}. We repeated our reproduction attempts three times under the same recorded conditions to confirm that the reported failure was consistently reproduced despite nondeterminism. We labelled a bug as reproduced only if all three executions satisfied the corresponding matching criterion. For crashes, we required an exact match with the reported error message. For silent bugs, we required numerical results to remain within 5\% of the reported failure value, per the existing criterion~\cite{shah2026imitation}.

For each bug, we recorded execution commands, outcomes, observations, and blocking reasons in \textit{reproduction.json}; investigation steps and adjustments in \textit{reproduction\_trajectory.md}; and runtime output in \textit{repro\_*.log}. These artifacts allow users to inspect the evidence behind each outcome.

Out of 770, we successfully reproduced 652 and failed to reproduce the remaining 118 within our three-hour budget and recorded execution conditions. Of these 118 reports, 46 required unavailable hardware, operating systems, builds, data, network resources, or credentials. Another 72 did not exhibit the reported behaviour on the recorded version of the codebase or were affected by version drift or nondeterminism. The documented reasons provide insights into practical barriers to debugging AI-based systems.

\subsection{Cleaning and Structural Validation} We normalized bug identifiers into three-digit strings and organized each bug's artifacts under \textit{bugs/<id>/} (Fig.~\ref{fig:collection-process}). Our supplied validator checks required files, directory structure, and schema compliance. It also verifies that the CSV index, JSON index, and directory tree contain the same identifiers and that shared fields agree.

\section{Internal Structure}

\bench stores fault metadata in two top-level index files and keeps each bug's artifacts in a separate directory~\cite{aifaultbench}. The \textit{index.csv} file provides one row per bug report and can be analyzed using tools such as Excel and pandas. The \textit{index.json} file contains records that share fields with \textit{index.csv} (e.g., bug identifier and repository). It also includes source revisions, execution commands, report paths, and a nested \textit{github} object. The \textit{github} object contains the bug metadata, comments, and available fix information.

Both index files identify each bug using the same \textit{bug\_id}, which maps to its directory under \textit{bugs/<bug\_id>/}. For example, the record with identifier \textit{001} corresponds to \textit{bugs/001/}. The \textit{bug\_report} field in the \textit{index.json} also provides the relative path to the report, such as \textit{bugs/001/bug\_report.txt}. Reports, scripts, and execution evidence are stored as separate files within each bug's directory. Table~\ref{tab:structure} distinguishes between bug indices and their storage locations. To inspect the evidence of a fault, one can navigate from the \textit{bug\_id} to the corresponding directory containing \textit{reproduction.json}, \textit{reproduction\_trajectory.md}, and the available logs.

\begin{table}[t]
\caption{Structure of \bench}
\label{tab:structure}
\centering
\footnotesize
\setlength{\tabcolsep}{4pt}
\renewcommand{\arraystretch}{1.15}
\begin{tabularx}{\columnwidth}{@{}lX@{}}
\toprule
\textbf{File} & \textbf{Contents} \\
\midrule

\multicolumn{2}{@{}l}{\textbf{Dataset indices}} \\
\addlinespace[2pt]
\textit{index.csv} &
Bug identifiers, system classes, repositories, reproduction status, and issue summaries. \\
\textit{index.json} &
Bug records with source revisions, report paths, execution commands, and issue details. \\

\midrule
\multicolumn{2}{@{}l}{\textbf{Per-bug files: report and metadata}} \\
\addlinespace[2pt]
\textit{bug\_report.txt} &
Original bug report. \\
\textit{github.json} &
Issue metadata and discussions. \\
\textit{manifest.json} &
Source and packaging metadata. \\

\addlinespace[5pt]
\multicolumn{2}{@{}l}{\textbf{Per-bug files: setup and execution}} \\
\addlinespace[2pt]
\textit{requirements.txt} &
Dependency requirements. \\
\textit{setup\_codebase.sh} &
Source retrieval script. \\
\textit{setup\_env.sh} &
Environment setup script. \\
\textit{repro.py}, \textit{run\_repro.sh} &
Reproduction trigger and execution script. \\

\addlinespace[5pt]
\multicolumn{2}{@{}l}{\textbf{Per-bug files: results and evidence}} \\
\addlinespace[2pt]
\textit{reproduction.json} &
Reproduction outcome and supporting evidence. \\
\textit{reproduction\_trajectory.md} &
Investigation record. \\
\textit{repro\_*.log} &
Execution output, where available. \\

\bottomrule
\end{tabularx}
\end{table}

Users can store experimental results in a separate CSV or JSON file with a \textit{bug\_id} field and join these results with the benchmark index. Additional logs and tool outputs can be stored in a separate results directory organized by the same identifiers. The hierarchical structure of our benchmark dataset provides a systematic way to explore faults and relevant artifacts while avoiding information overload.

\section{Access, Usage, and Demonstration}

\noindent\textit{a) Dataset Access and Reproduction Procedure:} The dataset is available through Zenodo,\footnote{\url{https://zenodo.org/records/21782307}} Hugging Face,\footnote{\url{https://huggingface.co/datasets/mehilshah/AIFaultBench}} and GitHub.\footnote{\url{https://github.com/mehilshah/AIFaultBench}} Zenodo provides the archival release, Hugging Face provides access to dataset records and metadata, and GitHub provides documentation, reproduction scripts, the validator, and an
analysis notebook.  
Reproduction requires Git, a POSIX-compatible shell, and the Python version required by the selected subject system. Additional requirements include accelerators, input data, network access, or credentials. Users should check the setup scripts and investigation record before preparing the environment.

A typical workflow involves filtering the fault index with appropriate options (e.g., \textit{reproducible=true}) and choosing the entries (bugs) of interest. For each selected bug, one can enter its directory and run \textit{setup\_codebase.sh} and \textit{setup\_env.sh} to retrieve the recorded source revision and install its dependencies, respectively. Then, the corresponding \textit{run\_repro.sh} can be executed, and the observed behaviour can be compared with that recorded in \textit{reproduction.json} and the execution logs. If the expected failure does not occur, \textit{reproduction\_trajectory.md} provides necessary guidance on failed preparation, blocked execution, or other unexpected scenarios.

\noindent\textit{b) Typical Uses and Target Users:} \bench can support various users---empirical researchers, AI tool developers, AI practitioners, and educators---in their tasks.

\textbf{Empirical researchers} can investigate how fault symptoms, causes, and reproduction barriers vary across different classes of AI systems. Using the links to fixes, they can also mine repair patterns and maintenance practices in AI-based software systems. When combined with a suitable conventional software dataset, \bench can support comparisons between AI and conventional software faults, revealing important insights. Such studies must account for differences in sampling and classification.

\textbf{Tool developers} can use the bug reports, affected source versions, and documented reproduction procedures to evaluate their debugging tools across different classes of AI systems. Potential applications include fault reproduction, localization, automated repair, and debugging support within development environments. The same benchmark can be used to contrast between multiple techniques. If calibrated carefully, it can also help detect regressions within the same AI-based software systems.

\textbf{AI practitioners} can use documented failures to identify relevant dependency, data, or hardware conditions when debugging their own systems. The setup scripts, execution logs, and investigation records provide examples of how a reported problem was examined. Practitioners can also adapt relevant triggers into tests for their systems, provided that the expected behavior and execution conditions apply. Across a set of connected bugs, these tests might enable integration or system-level testing and validation against known failure patterns.

\textbf{Educators} can use individual bug records as case studies in software testing, debugging, and AI system maintenance. Students can compare an original report with the execution evidence, examine environment requirements, and explain why a reproduction attempt succeeded or failed.

\noindent\textit{c) Showcase Walkthrough:} Our demonstration follows an existing bug from the index to its report, affected codebase, scripts, and recorded evidence. It not only compares the observed behaviour with the reported failure but also examines an unsuccessful attempt and its resulting artifacts. Thus, our examples show researchers how to inspect evidence relevant to an AI fault, tool developers how to select evaluation inputs, practitioners how to investigate failure conditions, and educators how to explain a documented bug investigation.

\section{Limitations}

\bench has a few limitations. First, it includes public GitHub issues and may not represent faults across all AI-based software systems. The selection criteria favor active Python repositories with established community interest and sufficient closed issues. As a result, the benchmark might not represent smaller, less active, private, or non-Python systems. Therefore, the reported fault distributions across classes and reproduction rates should be considered properties of this benchmark dataset, rather than the actual fault prevalence across all AI-based software. Second, fault reproduction depends on execution conditions; any changes to dependencies, services, or hardware can prevent users from observing a reported failure. Our recorded commit identifiers specify the intended source revision, but deleted repositories, unavailable packages, or rewritten history can still prevent environment reconstruction and, in turn, the future execution of the released reproduction scripts. Third, the reproduction evidence does not guarantee repair correctness. Repair evaluations must supplement the provided evidence with project-specific tests or additional behavioural checks where needed.

\section{Responsible Use}

The benchmark preserves public contributor information and issue discussions to maintain fault provenance. Users should retain the context of these discussions and avoid using contributor metadata to infer sensitive personal attributes. Original source code and reports remain subject to their applicable upstream licenses. Because reproduction installs historical dependencies and executes repository code, users should inspect the setup procedures and use disposable environments without sensitive files or credentials. Execution costs vary with hardware, dependencies, and external resources. We do not report a measured carbon footprint. Future resource profiles and additional reproduction runs could support cost estimation and assessment of changes in reproducibility.

\section{Conclusion}

\bench provides 770 real-world faults from 105 AI-based software systems across six classes of AI. It complements existing benchmarks focused on conventional software or specific classes of AI systems by bringing faults from multiple classes of AI systems into a common structure. Each fault connects to its report and affected codebase with structured metadata, reproduction procedures, and investigation evidence. Our dataset includes 652 reproduced faults and documented outcomes for the remaining 118 records.

It has strong potential to support research on fault characteristics and maintenance practices, evaluation of debugging techniques, and education through real-world examples. Availability through Zenodo, Hugging Face, and GitHub provides archival access, access to structured records, and supporting documentation and scripts. Together, these resources provide a practical basis for systematically understanding and debugging complex failures in AI-based software.

\balance
\bibliographystyle{IEEEtran}
\bibliography{main-full}
\end{document}